\RequirePackage{fix-cm}
\documentclass[smallextended]{svjour3}       
\smartqed  

\usepackage{amssymb}
\usepackage{amsmath}
\usepackage{physics}
\usepackage{algorithmic}

\usepackage{physics}
\usepackage{mathtools}

\usepackage{array}

\usepackage[caption = false, font = footnotesize, labelfont = sf, textfont = sf]{subfig}

\usepackage{floatrow}
\usepackage{tabularx}
\usepackage{multirow}
\usepackage{hhline}
\usepackage[table]{xcolor}

\usepackage[export]{adjustbox}
\usepackage{pgfplots}
\usepackage[T1]{fontenc}
\usepackage[latin9]{inputenc}
\usepackage{epsfig}
\usepackage{collcell}

\usepackage{pgf}
\pgfplotsset{compat=1.14}

\usepackage{subfig}

\usepackage{caption}
\DeclareGraphicsExtensions{.pdf,.jpeg,.png,.bmp,.jpg,.tif,.eps}
\newcolumntype{P}[1]{>{\centering\arraybackslash}p{#1}}
\newcolumntype{M}[1]{>{\centering\arraybackslash}m{#1}}
\newcolumntype{R}[1]{>{\raggedleft\arraybackslash}m{#1}}

\newcommand{\etal}{\textit{et al}.~}

\usepackage[nocompress]{cite}
\usepackage{stfloats}

\usepackage{hyperref}
\hypersetup{hidelinks,bookmarksopen=true} 

\usepackage{color}

\begin{document}

\title{Design of a Quantum-Repeater using Quantum-Circuits and Benchmarking its Performance on an IBM Quantum-Computer
}

\titlerunning{Design of a Quantum Repeater using Quantum Circuits}        

\author{Sowmitra Das         \and
        Md. Saifur Rahman   \and 
        Mahbub Majumdar
}

\authorrunning{Das et. al.} 

\institute{Sowmitra Das \at Dept. of Theoretical Physics, University of Dhaka, Dhaka, Bangladesh\\
School of Data and Sciences, BRAC University, Dhaka, Bangladesh \\
Tel.: +880-1747-221216\\
\email{sowmitra.das@bracu.ac.bd} 
\and
Md. Saifur Rahman \at  Dept. of Electrical and Electronic Engineering, \\Bangladesh University of Engineering and Technology, Dhaka, Bangladesh\\
  \email{saifur@eee.buet.ac.bd}
\and 
Mahbub Majumdar \at  School of Data and Sciences, BRAC University, Dhaka, Bangladesh \\
 \email{majumdar@bracu.ac.bd}
}

\date{Published in Springer Quantum Information Processing 20, 245 (2021)\\
Received: 8 December 2020 / Accepted: 12 July 2021\\
DOI: \url{https://doi.org/10.1007/s11128-021-03189-8}\\
Sharedit Link: \url{https://rdcu.be/czwI5}}

\maketitle

\begin{abstract}
Quantum communication relies on the existence of entanglement between two nodes of a network. However, due to its fragile nature, it is nearly impossible to establish entanglement at large distances through the direct transmission of qubits. Quantum repeaters have been proposed to solve this problem, which split-up the network to create small-scale entangled links and then connect them up to create the large-scale link. As researchers race to establish entanglement over larger and larger distances, it becomes essential to gauge the performance and robustness of the different protocols that have been proposed to design a quantum repeater, before deploying them in real life. Currently available noisy quantum computers are ideal for this task, as they can emulate the noisy environment in a quantum communication channel, and provide a measure for how the protocols will perform on real-life hardware. In this paper, we report the circuit-level implementation of the complete architecture of a quantum repeater, and benchmark this protocol on IBM's cloud quantum computer - IBMQ. Our experiments indicate a 26\% fidelity of shared bell-pairs for a complete on-chip quantum repeater with a yield of 49\%. We also compare these results with simulation data from IBM Qiskit. The results of our experiments provide a quantitative measure for the fidelity of entanglement that currently available repeaters can establish. In addition, the proposed circuit-implementation provides a robust benchmark for state-of-the-art quantum computing hardware.
\keywords{Quantum Communication \and Quantum Repeaters \and Entanglement Swapping \and Entanglement Purification \and Quantum Circuits \and IBMQ}
\end{abstract}

\section{Introduction}
Quantum communication is the method of transmitting information signals by exploiting the the principles of quantum mechanics\cite{bennett1984g, gisin2007quantum}. It enables novel communication paradigms such as quantum teleportation\cite{bennett1993teleporting}, superdense coding\cite{bennett1992quantum}, and unbreakable cryptography\cite{bennett1992communication, gisin2002quantum} - all of which have no classical counterpart. These protocols require the existence of entanglement between the transmitting and receiving parties. Photonic channels have proved to be a reliable medium for communicating classical signals over long distances. However, quantum entanglement is an extremely fragile resource, and, the smallest amount of noise (thermal or otherwise) in the environment can render them useless. The attenuation length of classical optical-fiber based photonic channels for entangled quantum signals is up to a hundred kilometers for deployed telecommunication fibers, and several hundred kilometers for ultra low-loss fibers in laboratory settings\cite{wengerowsky2019entanglement, wengerowsky2020passively, inagaki2013entanglement}. As a result, it is extremely difficult to transmit entangled photons over large distances preserving their fidelity. Classical repeaters tackle this issue of loss by simply amplifying, or by measuring and regenerating, the input signal. However, the no-cloning theorem forbids the amplification of quantum signals\cite{wootters1982single, dieks1982communication}, and, decoherence does not allow us to measure quantum systems without destroying their information content\cite{park1970concept}. This poses a big problem for transmitting these signals reliably.

Quantum repeaters have been proposed to solve this problem\cite{briegel1998quantum}. The idea behind this mechanism is to divide the entire link into many small segments - the length of each segment being less than the attenuation length of the channel. Entanglement is established between the endpoints of each of these smaller links by direct transmission of photons. Then, by using the entanglement swapping protocol, all of these smaller links are connected up to establish the large-scale link. At each successive step of this process, there might be loss of fidelity due to noise or imperfections in the operating hardware. Therefore, at each step, the entangled links are "purified" to increase the fidelity of entanglement between the nodes\cite{bennett1996purification}. By repeating this protocol a sufficient number of times, we can theoretically establish a large-scale entangled link of an arbitrarily high fidelity. 

These ideas are already being tested in the field as researchers around the globe race to establish entanglement over larger and larger distances, spanning thousands of kilometres. Significant progress has been made in recent years\cite{Ruihong_2019, li2019experimental, rozpkedek2019near, borregaard2020one}. However, real-life deployment of entanglement still remains a difficult problem, due to the extremely noisy and unpredictable nature of the quantum channel, the sensitivity of quantum signals to external influence, and, the stringent conditions and high-precision instruments required to design the systems. As we enter the Noisy Intermediate-Scale Quantum (NISQ) era, quantum communication technologies are poised to disrupt and revolutionize the entire communication infrastructure\cite{Preskill2018quantumcomputingin}. Therefore, it is more important than ever, that we have the ability to predict how these protocols and systems will behave in real quantum environments. This ability will enable us to test the robustness of our algorithms before committing resources to deploy them in real life. Unfortunately, it is exceedingly difficult to model a quantum channel by using classical resources or hardware. This classical approach requires enormous amounts of computational resources, and, for larger systems, can fall short of exactly capturing the quantum-mechanical behaviour of the channel altogether - even if it is possible at all. Therefore, instead of simulating these channels and protocols classically, the best way forward would be to use actual quantum-mechanical hardware to emulate their effects\cite{feynman1982simulating}. 

In light of the above discussion, it is insightful to see the performance of these protocols on current NISQ devices. One of these devices is IBM's cloud quantum computer - IBMQ \cite{cross2018ibm}, which can be openly accessed by the general public and researchers using IBM's Quantum Information Science toolKIT (QISKIT)\cite{aleksandrowicz2019qiskit} API.
The noisy hardware of IBMQ can emulate the conditions encountered in real-life quantum systems, particularly a quantum communication channel. Thus, it is a perfect candidate to gauge the performance of current quantum communication protocols, and can be a useful guide for further research to make these protocols as robust as possible to be applicable in real-life scenarios. With this end in view, we move to implement the complete end-to-end architecture of a Quantum Repeater on an IBM Quantum Computer. 

A similar demonstration of this scheme has been given by Behera \etal in \cite{behera2019demonstration}. However, they have performed limited experiments in this regard, showing the results of entanglement swapping on 2 pairs of qubits only. In quantum-repeaters, entanglement-swapping has to be performed on multiple pairs of qubits successively in a nested fashion. As a result, hardware errors might accumulate in qubits impacting their fidelity. Therefore, performing the entanglement - swapping protocol only once, most likely would not provide representative results of a real quantum repeater. In addition, they have also introduced a  quantum error-correction code to purify the entangled-qubits by using Controlled-NOT (CNOT) operations on them. This approach is also not suitable for large-scale quantum networks since CNOT, being a local-operation, cannot be performed on qubits separated by a large distance in space. In addition, they report that, the error-correction code can only correct 2 specific types of errors - namely, bit flip and phase-change - and that too, only if the entangled qubits start out are in a pure-state. For more general types of errors in a quantum-communication channel, and for qubits starting out in a mixed-state, their code fails to increase the fidelity of entanglement. For communication networks, this can be mitigated by condensing multiple low-fidelity entangled pairs into one high-fidelity pair by using an entanglement-distillation protocol. We demonstrate this general approach of building a quantum-repeater in our experiments.

Finally, we not only demonstrate a complete circuit-level implementation of a Quantum Repeater, but the result of our experiments also sheds light on the performance and accuracy of the current quantum hardware. Zhukov \etal \cite{zhukov2019quantum} have proposed that quantum communication protocols can serve as a deep benchmark for quantum computers. In contrast to generic protocols, like superdense coding and quantum teleportation, which require only a few gate operations, the complete architecture of a Quantum Repeater uses multiple sub-protocols and stresses every physical aspect of the hardware platform on which they are run. As a result, the quantum-repeater protocol can serve as a much more revealing benchmark compared to other protocols. This will offer us several metrics reflecting the real-life performance of Quantum Repeaters, and, help us evaluate the hardware limitations of current state-of-the-art noisy quantum computers.

The following sections of this paper are organised as follows. Section 2 introduces the basics of Quantum Circuits and the process of designing them for Quantum Computing hardware. The basic protocols of Quantum Communication, for which entanglement-distribution is a key step, are shown next in Section 3. Then, we move on to designing the fundamental Quantum Repeater architecture in Section 4, which includes the entanglement-distribution, entanglement swapping and entanglement purification protocols. In section 5, we report the results of our experiments on the IBMQ platform, as well as simulation results with noise models for the purpose of comparison. Finally, we analyze the results in Section 6 and explore the future directions of our work with discussions on how to improve the yield and accuracy of the protocol even further.  

\section{Quantum Circuits}
\subsection{Qubits and 1-Qubit Gates} 
The fundamental building-block of a Quantum Circuit is the `Qubit'. It is a two-dimensional quantum system with orthonormal basis-states $\ket{0}$ and $\ket{1}$ (known as the Computational Basis) \cite{deutsch1989quantum}. The state $\ket{\psi}$ of a system can be any superposition of the basis states, which can be expressed as, 
\[\ket{\psi} = \alpha\ket{0} + \beta\ket{1} = \mqty[\alpha\\\beta] \]
where, $\alpha$ and $\beta$ are complex numbers such that $\abs{\alpha}^2 + \abs{\beta}^2 = 1$. The last equality asserts that, the states of a qubit are essentially vectors in a complex Hilbert Space, and, the whole machinery of linear algebra may be used in their manipulations. The orthonormal basis states might, for example, be the ground state $\ket{g}$ and an excited state $\ket{e}$ of a matter system or the horizontally polarised state $\ket{H}$ and vertically polarised state $\ket{V}$  of a single-photon system. In the context of a Quantum-Computing system, a collection of qubits form a \emph{Quantum Register}, whereas, a collection of classical bits form a \emph{Classical Register}\cite{nielsen_chuang_2010}. 

\begin{figure}[h]
\centering
\includegraphics[scale=0.6]{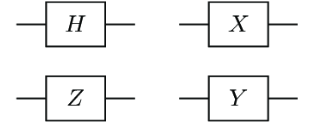}
\caption{Schematic Symbols for Quantum Logic Gates (in clockwise order) - Hadamard Gate, Pauli X (Not) Gate, Pauli Y Gate, Pauli Z Gate.}
\label{fig_gates}
\end{figure}

Qubits may be manipulated by using \emph{Operators}, which are represented as matrices. Operators which preserve the normalization of a state are called \emph{Unitary Operators}. At the circuit-level, unitary operators are implemented by using Quantum Logic-Gates, or simply, Quantum-Gates. The prototypical quantum gates are the Pauli Gates $X, Y, Z$, the Hadamard Gate $H$ and the Phase-Shift Gate $R_{\theta}$\cite{barenco1995elementary}. Their matrix-representations are given below and circuit symbols are shown in Fig. \ref{fig_gates}.
\[X = \mqty[0 & 1 \\ 1 & 0] \qquad Y = \mqty[0 & -i \\ i & 0] \qquad Z = \mqty[1 & 0 \\ 0 & -1]\]
\[H = \dfrac{1}{\sqrt{2}}\mqty[1 & 1 \\ 1 &-1] \qquad R_{\theta} = \mqty[1 & 0 \\ 0 & e^{i\theta}]\]

\subsection{Ebits and 2-Qubit Gates}
Two qubits $A$ and $B$ may exist in a tensor product-state $\ket{\psi}_A\otimes\ket{\phi}_B$, or written simply as $\ket{\psi}_A\ket{\phi}_B$, so that each qubit may be assigned an individual state vector $\ket{\psi}$ or $\ket{\phi}$. However, the remarkable aspect of Quantum Mechanics is the existence of \emph{Entanglement} between 2 qubits, where, the qubits may be in a superposition of product-states, but, separate state-vectors can not be assigned to them. The prototypical entangled states of a two-qubit system $AB$ are the maximally entangled Bell-States, $\ket{\Phi^{\pm}}$ and $\ket{\Psi^{\pm}}$, which are expressed in terms of the computational basis states as follows:
\[\ket{\Phi^{\pm}}_{AB} = \left(\ket{0}_A\ket{0}_B \pm \ket{1}_A\ket{1}_B\right)/\sqrt{2}\]
\[\ket{\Psi^{\pm}}_{AB} = \left(\ket{0}_A\ket{1}_B \pm \ket{1}_A\ket{0}_B\right)/\sqrt{2}\]

Two qubits in a Bell-State are also known as a Bell-Pair \cite{bell1964einstein} or an EPR (Einstein-Podolsky-Rosen) Pair\cite{einstein1935can}. 

The Bell-States may be prepared by using the two-qubit Controlled-NOT (CNOT) Gate. The CNOT Gate flips the state of the target qubit ($\ket{0}$ to $\ket{1}$, or, $\ket{1}$ to $\ket{0}$) if the state of the control qubit is $\ket{1}$, and, it does nothing if the state of the control qubit is $\ket{0}$. If the control qubit is in a superposition state, the control and target qubits become entangled at the output of the gate. Therefore, the CNOT Gate is a mechanism to entangle two qubits. Two qubits which are entangled in this way are called `Entangled Bits' or `Ebits' \cite{schumacher2010quantum, nielsen_chuang_2010}. 

The map of the CNOT Gate is shown as follows:
\[\ket{a}_{\textsf{control}}\ket{b}_{\textsf{target}}\xrightarrow{\text{CNOT}}\ket{a}_{\textsf{control}}\ket{a\oplus b}_{\textsf{target}}\]
where, $a,b\in \{0, 1\}$, and, $\oplus$ is the XOR operation. \\


Another example of a two-qubit gate is the SWAP gate, which exchanges the contents of the target and control qubit. 
\[\ket{\psi}_{\textsf{control}}\ket{\phi}_{\textsf{target}}\xrightarrow{\text{SWAP}}\ket{\phi}_{\textsf{control}}\ket{\psi}_{\textsf{target}}\]

The circuit symbols of the CNOT Gate and SWAP Gate are shown in Fig. \ref{fig_2qubit_gates} below.

\begin{figure}[!h]
\centering
\includegraphics[scale=0.8]{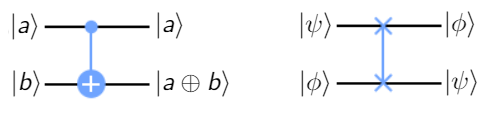}
\caption{2-Qubit Quantum Gates. (Left) Controlled-NOT (CNOT) Gate. $\ket{a}$ and $\ket{b}$ are the control and target qubits respectively. Here, $a,b \in \{0,1\}$  (Right) SWAP Gate.}
\label{fig_2qubit_gates}
\end{figure}

The Hadamard Gate, Phase-Shift Gate, Pauli Gates and the CNOT Gate form a \emph{Universal} Gate Set using which any Quantum Circuit may be constructed. 

\subsection{Measurement and Conventions} 
Finally, after completing a Quantum Computation task, a qubit may be measured out in the computational basis. This causes the state of the qubit to collapse to one of the basis states $\ket{0}$ or $\ket{1}$. The value obtained may be stored in a classical register so that they may be used for further processing.

Conventions of Quantum-Computing dictate that all qubits must start out in the state $\ket{0}$ and can only be measured in the computational basis. Other states have to be prepared by applying necessary Unitary Operations, and measurement in other bases may be performed by using suitable gates before the measurement operation \cite{nielsen_chuang_2010}.

\begin{figure}[!h]
\centering
\includegraphics[scale=0.55]{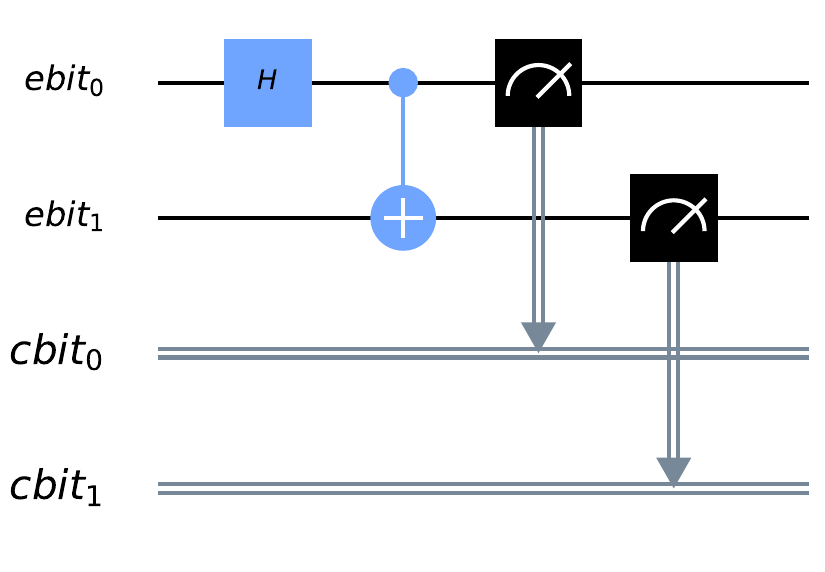}
\caption{Quantum Circuit for preparing the Bell-State $\ket{\Phi^{+}}$, measuring it, and storing the values in a Classical Register.}
\label{fig_bell}
\end{figure}

A quantum circuit for preparing the Bell-State $\ket{\Phi^{+}}$ and measuring it in the computational-basis illustrates all the operations outlined in this section. (Fig. \ref{fig_bell})

\section{Quantum Communication}
The main role of quantum communication is to transmit quantum signals (i.e, quantum states) over large distances  from a party Alice to a distant party Bob\cite{bennett1984g, gisin2007quantum}. Ideally, Alice's state $\ket{\psi}_A$ should be transferred to Bob without any change in its contents. Therefore, the ideal quantum channel (for a 1-qubit state) could be described by the identity map 
\[I^{A\rightarrow B} = \ket{0}_B\bra{0}_A + \ket{1}_B\bra{1}_A\]
so that, $I^{A\rightarrow B}\ket{\psi}_A = \ket{\psi}_B$. Hence, one might suppose that the goal of performing quantum communication is to give Alice and Bob a device to work as the identity map $I^{A\rightarrow B}$\cite{munro_inside_2015}. However, interestingly, this is not the only solution. In particular, quantum communication is also possible if the distant parties, Alice and Bob, share an entangled Bell state $\ket{\Phi^{+}}_{AB}$. In this section, we illustrate this approach through providing representative quantum communication operations, i.e., Quantum Teleportation \cite{bennett1993teleporting}, Superdense Coding\cite{bennett1992communication}, and Quantum Key Distribution\cite{ekert1991quantum,bennett1984g}. As a result, the Bell state is regarded as a resource for quantum communication. \cite{munro_inside_2015}

\subsection{Quantum Teleportation}
Quantum teleportation \cite{bennett1993teleporting} is an important primitive of quantum communication operations. If Alice and Bob share an entangled Bell-state between them, then, by using local quantum operations on their respective qubits and classical communication, Alice can transmit any 1-qubit state to Bob. This scheme is known as LOCC (Local Operation + Classical Communication)\cite{peres1991optimal}. 

The scheme begins with Alice and Bob having access to the corresponding qubits of a Bell-State $\ket{\Phi}_{AB}$. Alice has another qubit in the state -  $$\ket{\psi}_{A'}= \alpha\ket{0}_{A'} + \beta\ket{1}_{A'}$$ Then, the state of the combined system $A'AB$ is 
\begin{align*}
    \ket{S}_{A'AB}  &= \ket{\psi}_{A'}\otimes\ket{\Phi}_{AB} \\
                    &= \frac{1}{2}\ket{\Phi^+}_{A'A}\otimes\ket{\psi_1}_B +\frac{1}{2}\ket{\Phi^-}_{A'A}\otimes\ket{\psi_2}_B\\ 
                    \;&+\frac{1}{2}\ket{\Psi^+}_{A'A}\otimes\ket{\psi_3}_B + \frac{1}{2}\ket{\Psi^-}_{A'A}\otimes\ket{\psi_4}_B 
\end{align*}

where, $\ket{\psi_1}=\alpha\ket{0} + \beta\ket{1}$, $\ket{\psi_2}=\alpha\ket{0} - \beta\ket{1}$, $\ket{\psi_3}=\alpha\ket{1} + \beta\ket{0}$, $\ket{\psi_4}=\alpha\ket{1} - \beta\ket{0}$.
We can see that the state $\ket{\psi_1}$ is the same as the original state $\ket{\psi}$. So, if Alice performs a Bell-Basis measurement on her qubits $A'$ and $A$, and, gets the result $\ket{\Phi^+}$, then, we can be sure that the state $\ket{\psi}$ has been transferred to Bob unchanged. 

However, the result of of Alice's measurement may also be $\ket{\Phi^-}$, $\ket{\Psi^+}$ or $\ket{\Psi^-}$, in which case the state of Bob's qubit will be $\ket{\psi_2}$, $\ket{\psi_3}$ and $\ket{\psi_4}$ respectively. These states are different from the original state $\ket{\psi}$. Upon closer observation, we can see that $\ket{\psi_2}$ differs from $\ket{\psi}$ by a phase-flip, $\ket{\psi_3}$ by a bit-flip and $\ket{\psi_4}$ by both a phase-flip and a bit-flip. Hence, if Alice communicates the result of her measurement to Bob (which requires only classical communication) Bob can perform the corresponding correction operation - $I$ (nothing), $Z$ (phase-flip), $X$ (bit-flip), or $ZX$ (phase-and-bit-flip) - on his qubit, and, he would have Alice's original state $\ket{\psi}$ at his disposal. Thus, Alice can transfer the state of her qubit $A'$ just by using the Bell-State $\ket{\Phi^+}$ and classical communication\cite{schumacher2010quantum}. 

Since quantum teleportation requires classical communication from Alice to Bob, Bob's system B should work as a quantum memory to keep the quantum state until at least the end of the classical communication. This is important, as it inherently implies that quantum teleportation requires the use of a quantum memory\cite{munro_inside_2015}. 


\subsection{Superdense Coding}
The mathematical formalism of Superdense Coding\cite{bennett1992communication} is similar to that of the Teleportation Protocol. In this, we use the fact that, any local operation by Alice on her qubit of the shared Bell-Pair $\ket{\Phi^{+}}_{AB}$ from the set $\{I_A, Z_A, X_A, Z_AX_A\}$ can change the state of the total Bell-Pair system to the states $\{\ket{\Phi^{+}}_{AB}, \ket{\Phi^{-}}_{AB}, \ket{\Psi^{+}}_{AB}, \ket{\Psi^{-}}_{AB}\}$. Now, if Alice sends her part of the Bell-Pair to Bob via a quantum channel, and Bob performs a Bell-basis measurement on both the qubits, he can find out which operation Alice performed. Since there are 4 different operations, Alice has transferred the equivalent of 2 bits of information by transmitting a single qubit. This has remarkable possibilities for high-rate data-communication, since the Bell-pairs may be shared beforehand, and the channel-bandwidth can be utilized only during the actual time of communication. 


\subsection{Quantum Key Distribution}
Suppose Alice and Bob possess qubits $A$ and $B$ of the 2-qubit system $AB$ in the Bell-State $\ket{\Phi^{+}}_{AB}$, and they measure each of their qubits in the computational basis. The Bell state, being a pure state, is not entangled with any other qubits. Therefore, the result of the measurement does not leave any trace on any other part of the environment from which they can be predicted. As a result, their bits are perfectly secure. In addition, from the definition of the Bell state $\ket{\Phi^{+}}_{AB} = (\ket{00}_{AB} + \ket{11}_{AB})/\sqrt{2}$, the computational basis measurement outcomes always result in both 0's or both 1's, with each occurring randomly 50\% of the time. Hence, invoking the one-time pad protocol, Alice and Bob can share a secret bit using the qubits $A$ and $B$ in an information-theoretically secure manner\cite{ekert1991quantum, bennett1992quantum}. Therefore, the Bell-Pair can be considered as a cryptographic resource. 



\section{Quantum Repeater}
The quantum communication tasks outlined in the previous section must start with the distribution of entangled Bell-Pairs\cite{bell1964einstein} or EPR-Pairs\cite{einstein1935can} between 2 parties. These are maximally entangled 2-qubit states of the form - 
\begin{align}
    \ket{\Phi^{\pm}}_{AB} = \left(\ket{0}_A\ket{0}_B \pm \ket{1}_A\ket{1}_B\right)/\sqrt{2}\\
    \ket{\Psi^{\pm}}_{AB} = \left(\ket{0}_A\ket{1}_B \pm \ket{1}_A\ket{0}_B\right)/\sqrt{2}
\end{align}
This may be achieved by distributing entangled photons via traditional broadband optical fiber communication links. However, it should be noted that, the fidelity of entanglement decreases with the length ($L$) of the fiber exponentially as $e^{-L/L_0}$, where $L_0$ is the attenuation-length of the channel. This necessitates the use of quantum repeaters to allow long-distance communication with finite resources and reasonable rates. In a Quantum Repeater, three primary operations are required to create long-range Bell states \cite{munro_inside_2015}. These are:
\begin{enumerate}
    \item Entanglement Distribution: Creating entangled links between network nodes through the direct transfer of photons.
    \item Entanglement Purification: Creating a high-fidelity entangled state from several low-quality ones. 
    \item Entanglement Swapping: Connecting the entangled links of adjacent nodes using Bell-basis measurement to create long-range entanglement.
\end{enumerate}

Since direct transfer of photons needs to be done only between adjacent repeater nodes, and not across the entire long-range link, success probability for generating the entangled link depends only on the distance between adjacent nodes. This can be reduced arbitrarily to a small value, by dividing the long-range link into any required number of segments. 


\subsection{Quantum Purification}
A significant problem of the Bell states generated from an entanglement distribution scheme between the two remote nodes is that they are not perfect. While losses can be mitigated by repeating the scheme many times, other errors will occur in such systems. If the entangled states are stored in matter qubits of a Quantum Memory, they become highly prone to dephasing, i.e, the relative phase between the qubits changes spontaneously. Furthermore, there may be imperfections in the hardware used for state-preparation and measurement. These errors cannot generally be overcome by repetition, and thus, decrease the fidelity of the entangled link.

The fidelity $F$, of a system with density matrix $\rho$, which represents how close it is to a state with density matrix $\sigma$, is defined as,
\begin{equation}
    F(\rho, \sigma) = \left(\Tr\sqrt{\sqrt{\rho}\sigma\sqrt{\rho}}\right)^2
\end{equation}

Since, we are comparing the fidelity of our link to the state $\ket{\Phi^{+}}$, errors associated with imperfect local operations will decrease the fidelity of this state, inducing errors corresponding to the other three Bell-state elements. This is likely to lead to a maximally mixed state of the form:
\begin{align}
\rho_w =\; & F\ketbra{\Phi^{+}}{\Phi^{+}} +\nonumber\\
&\dfrac{1-F}{3}\left(\ketbra{\Phi^{-}}{\Phi^{-}} + \ketbra{\Psi^{-}}{\Psi^{+}} + \ketbra{\Psi^{-}}{\Psi^{-}}\right)
\end{align}
which is known as the Werner state \cite{werner1989quantum}.\\


\subsubsection{Bennett's Protocol}
\begin{figure}[!h]
\centering
\includegraphics[scale=0.65]{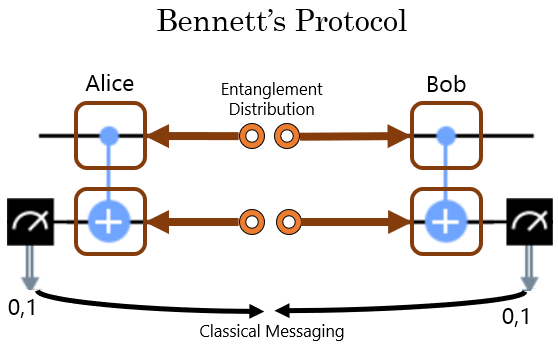}
\caption{Schematic illustration of an entanglement purification scheme using two imperfect Bell pairs and local operations including CNOT gates and projective measurements.}
\label{fig_recurrence}
\end{figure}

The decrease in the quality of entanglement means information present in the state has been lost. In general, it is not possible to recover quantum information without measurement, and thus, destroying the state in the process. However, since we are trying to generate a state that is known beforehand, we can distill a Bell state with higher fidelity from multiple imperfect copies of it by a process known as Quantum Purification \cite{bennett1996purification, bennett1996mixed, jiang2009quantum}. The original purification scheme was proposed by Bennett \etal \cite{bennett1996purification}  and is depicted in Fig. \ref{fig_recurrence}. 

It assumes that two copies of the Bell-state have already been established between repeater nodes (which may be of low fidelity). At each node, both the parties apply a CNOT operation between the two qubits keeping the corresponding qubits of each pair as control and target, respectively. The target qubits are then measured out in the computational basis \{$\ket{0}, \ket{1}$\}. Finally, the measurement-results are transmitted over a classical channel between the nodes. 

The unmeasured state of the control-qubits is kept only if the measurement results of the target-qubits agree (i.e, both are zero $(0,0)$ or one $(1,1)$). In such a case the purification is successful. The resulting state of the unmeasured qubit will be of a higher fidelity so long as the initial fidelity of both the pairs were greater than 50\% and our local operations (CNOT and the projective measurement) are accurate enough. If the measurement-results are not the same (i.e, $(0,1)$ or $(1,0)$), the purification protocol has failed, and one needs to start over again with fresh entangled states. This makes the purification protocol inherently probabilistic in nature, but it is heralded. Each party knows whether or not it was successful, but, only after the classical measurement-results have been exchanged between the nodes. This is likely to be a significant performance bottleneck. 

After a round of purification is performed, the higher-fidelity states may be used again to increase the fidelity even further. In this scheme, quantum purification can be performed using a `Recurrence Method' \cite{bennett1996purification}. However, if the purification fails during any step, the entire process has to be started again with a set of new pairs. 

This protocol can also be applied if the fidelity of the two Bell-Pairs are not equal\cite{briegel1998quantum,van2008system}, in which case the fidelity of the resulting state will be more than that of both the starting states. However, this does result in a relatively lower increase of fidelity per round than the case with equal fidelity states. 

\subsubsection{Deutsch's Protocol}
Bennett's protocol suffers from 2 major drawbacks. For it to work, firstly, the initial state must be of the Werner form. Secondly, it takes many rounds of purification to obtain a Werner state with a fidelity above 99\% when one starts with a set of low fidelity pairs (e.g., $F$ = 85\%). Deutsch \etal \cite{deutsch1996quantum} addressed these issues by modifying Bennett's protocol. 

The state of a two-level quantum system may be represented as a unit-vector in a 3-dimensional space. This is called the Bloch Vector of the state, and, the sphere on which it resides is called the Bloch Sphere. The unitary operation $R_x(\theta)$ represents rotating the Bloch Vector with respect to the $x$-axis by an angle $\theta$. In matrix notation, it is expressed as, 
\begin{equation}
    R_x(\theta) = \mqty[\;\;\cos{(\theta/2)} & -i\sin{(\theta/2)} \\ -i\sin{(\theta/2)} & \;\;\cos{(\theta/2)}]
\end{equation}

Deutsch et al. proposed that, before applying the CNOT Gate, Alice should perform a rotation $R_x(\pi/2)$ on her qubits, and, Bob should perform the inverse rotation $R_x(-\pi/2)$. All the other operations may be performed as in Benett's Algorithm. This procedure results in a theoretical increase in fidelity of about 100 times more than that of Bennett's. Moreover, the initial states of the Bell-Pairs need not be of the Werner form. \\



\subsubsection{Multi-Qubit Entanglement Purification}

\begin{figure}[!h]
\centering
\includegraphics[scale=0.6]{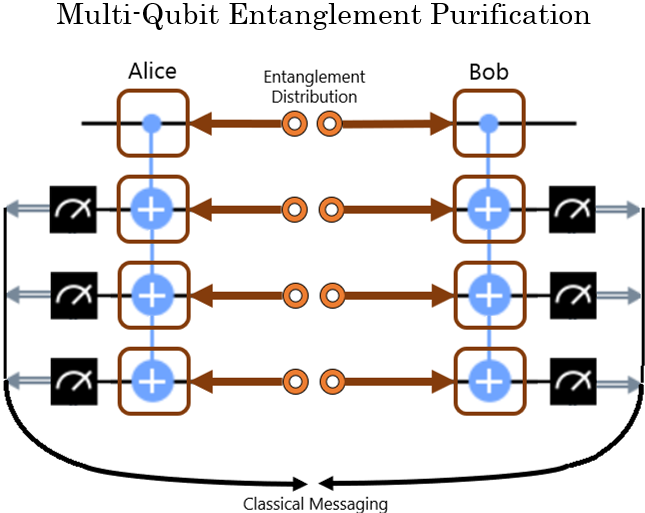}
\caption{ Schematic illustration of a generalised entanglement purification scheme using $n$ imperfect Bell pairs, local operations and classical communication.}
\label{fig_multi}
\end{figure}

In fact, Deutsch's Protocol can be further generalized. Instead of running the purification algorithm on 2 pairs, we can apply it on multiple pairs at the same time \cite{aschauer2005quantum, bennett1996mixed}. For instance, in the absence of measurement and gate-errors, the 5-qubit variant can theoretically purify 5 imperfect pairs with a fidelity of 0.85 into one with a fidelity above 99\% in a single round with a success probability of 0.44 \cite{munro_inside_2015}. Significantly more resources and communication time are required if one uses the recurrence method or entanglement pumping (refer to Fig. \ref{fig_recurrence}). A schematic for extending the purification protocol to multiple qubits is shown in Fig. \ref{fig_multi}.

\subsection{Entanglement Swapping}
Using entanglement distribution followed by quantum purification, we have a mechanism to generate a high-fidelity Bell-State between the adjacent repeater nodes. Now, we need a mechanism to connect the individual links together to form a long-range entangled link. This can be achieved with a protocol known as Entanglement Swapping \cite{bennett1993teleporting, briegel1998quantum}. 

\begin{figure*}[!h]
\centering
\captionsetup{justification=centering}
\includegraphics[scale=0.7]{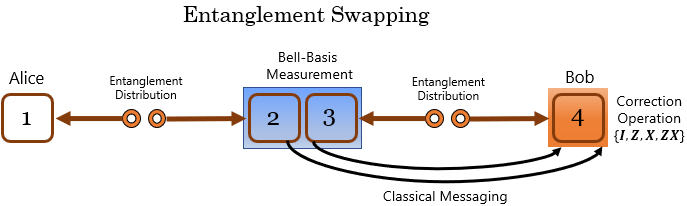}
\caption{Schematic of the Entanglement-Swapping Protocol.}
\label{fig_enswap2}
\end{figure*}

Consider that we have two Bell-Pairs in the combined state $ \ket{\Phi^{+}}_{12} \otimes \ket{\Phi^{+}}_{34} $, where the labels 1, 2, 3, 4 indicate the locations of the qubits - of which nodes $(1,2)$, $(2,3)$ and $(3,4)$ are adjacent (Fig. \ref{fig_enswap2}). When a Bell-state measurement between qubits 2 and 3 is performed, it projects qubits 1 and 4 into the state $\ket{\Phi^{+}}_{14}$ up to a Pauli correction operation $\{I, Z, X, Y \}$ depending on the result of the measurement. This result needs to be sent to qubit 4 (or qubit 1, but not both) so that the correction operation can be performed. This is essentially equivalent to the Teleportation protocol, but in this case, the qubit whose state is to be transferred is entangled with another qubit. 

The previous discussion assumed ideal Bell states after distribution and error-free operations for state-preparation. However, because of channel noise and imperfection of local devices, we will instead have mixed states. Modelling these as a Werner state $\rho_w$ with fidelity $F$, the resulting state after the Bell-state measurement and correction operations (which are assumed ideal) can be shown to be also a Werner state $\rho _{w_{14}}$ with fidelity $F'=F^2 + (1-F)^2/3$ \cite{briegel1998quantum}. This clearly shows that the fidelity of the longer-range state has decreased compared with the fidelity of the two initial entangled links. In fact, with a good approximation, $F' \approx F^2$ for $F \approx 1$. If one is performing entanglement swapping on multiple links (say $n$ links), the resulting fidelity will drop to $F' \approx F^n$. For error-prone Bell-State measurement and correction operations, the drop in fidelity is even sharper\cite{briegel1998quantum}. This implies that purification has to be performed on longer-range links after carrying out the swapping protocol. 

\subsection{Complete Architecture}
\begin{figure*}[!h]
\centering
\includegraphics[scale=0.7]{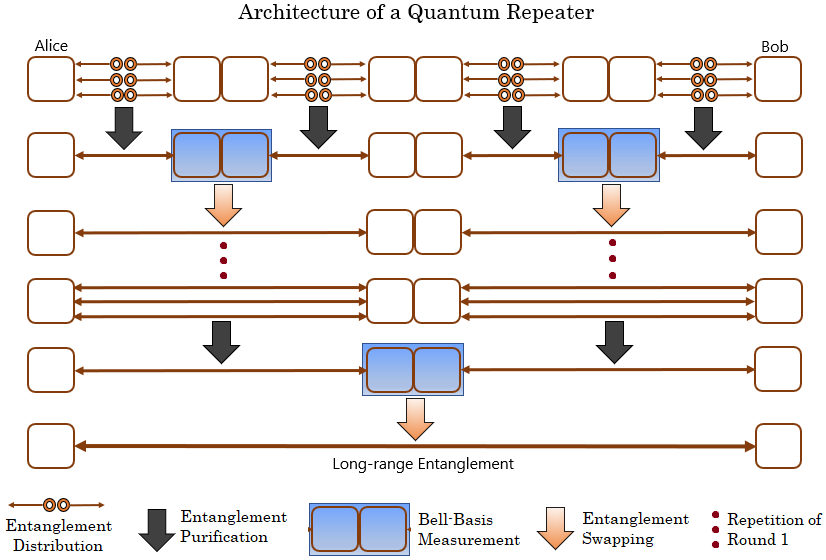}
\caption{The Quantum Repeater scheme for generating long-range entanglement: It begins with splitting the entire link into a number of segments and placing repeater stations at these nodes. Multiple entangled pairs are then generated between adjacent nodes. These shorter links are then purified and entanglement swapping is performed to create a link twice as long as the original one. These new links are then purified and entanglement swapping is performed again to create a link four times as long. This continues until entanglement is generated between the end repeater nodes (Alice and Bob).}
\label{fig_repfull}
\end{figure*}

Now that we have described all the components that go into a Quantum Repeater, we can illustrate how they are put together by referring to Fig. \ref{fig_repfull} and the steps of how a  Quantum Repeater operates. 

Step 1 - First, a number of Bell-Pairs are created between adjacent repeater nodes through entanglement distribution. After generating enough of them, entanglement purification is performed if necessary (either once or a number of times) to increase the fidelity of the link. Two neighbouring high-fidelity links are then connected by using the entanglement swapping protocol to generate a link twice as long as the original one.

Step 2 - Next, quantum purification is performed again on the longer links generated in Step 1. This is again followed by entanglement swapping to create even longer links. 
In this way, steps 1 and 2 are repeated until an entangled-link of required fidelity is generated between Alice and Bob. If the purification or entanglement swapping fails at any step, we must start over that part again from Step 1. After the entire pipeline of operations is complete, a robust and reliable entangled link will be established between the two parties\cite{briegel1998quantum}. 

The complete architecture of a Quantum Repeater combining all the components mentioned above is illustrated in Fig. \ref{fig_repfull}. Given that the communication link between Alice and Bob is divided into $n$ segments, a Quantum Repeater can establish long-range entanglement between them after at least $\left\lceil{\log_2n}\right\rceil$ number of rounds. 


\section{Experiments and Results}
All of the quantum circuits in the experiments were designed using IBM's open-source SDK -- Quantum Information Science toolKIT (QISKIT) in Python. 
The circuits were run on IBMQ-16-Melbourne - a real quantum computing device with 15 superconducting qubits, through back-end access via the cloud. \\
After performing the experiments, they were error-corrected using QISKIT's \textbf{Ignis} library and its Error-Mitigation protocols to remove the effects of measurement-errors in the results. 
In addition, the circuits were also simulated natively using QISKIT's `QASM Simulator' with a noise model from QISKIT's \textbf{Aer} library that mimics the device-noise of IBMQ-16-Melbourne. The simulation results provide a reference point to which the device results can be compared. 

All relevant Python codes and detailed simulation data can be found at {\color{red}{\url{https://github.com/SowmitraDas/Quantum-Repeater-using-Quantum-Circuits.git}}}.

\subsection{Channel-Length Simulation}
\begin{figure}[!h]
    \centering
    \includegraphics[scale=0.45]{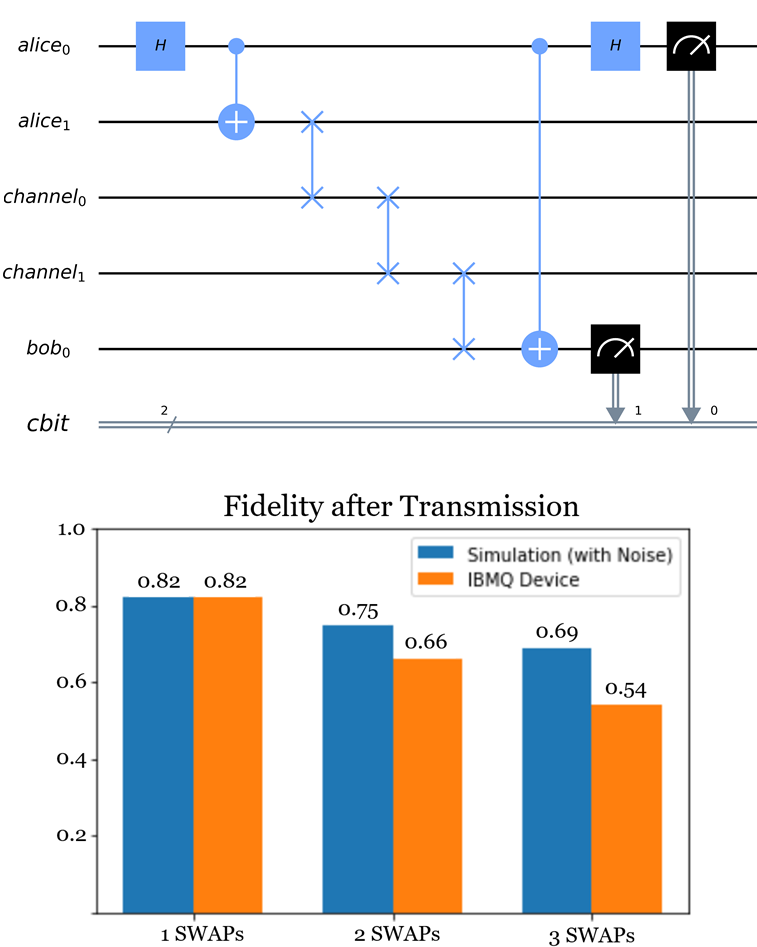}
    \caption{(Top) Quantum Circuit to simulate the effects of transmitting a qubit through a quantum channel (number of SWAP gates emulates channel-length). (Bottom) Effect of Channel-length on the fidelity of entanglement.}%
    \label{fig_channel}%
\end{figure}

First, the effect of channel-length on entanglement distribution was examined. The circuit used for this purpose is shown in Fig. \ref{fig_channel} (Top). 

Here, a Hadamard Gate ($H$) and CNOT gate are first used to create the Bell-state $\ket{\Phi^{+}}$. Transmitting a qubit of the Bell-Pair directly to a quantum channel is emulated by using the SWAP gates. The number of SWAP gates applied emulates the length of the channel. Finally, a Bell-basis measurement is performed to check the entanglement-fidelity of the transmitted state. The effect of the number of SWAP gates on the fidelity of entanglement is shown in Fig. \ref{fig_channel} (Bottom).


It is worth noting from the results that, just after 3 SWAP Gates, the fidelity of the Bell-pair falls below 50\%, making it unusable for further processing. Although a large portion of the readout errors are mitigated by QISKIT Ignis, on-chip errors for gate operations are still too error-prone and cause a large drop in fidelity. Thus, the number of SWAP operations is representative of the number of consecutive operations that can be performed faithfully on current quantum computers.  


\subsection{Quantum Purification}
\begin{figure}[!h]
\includegraphics[scale=0.45]{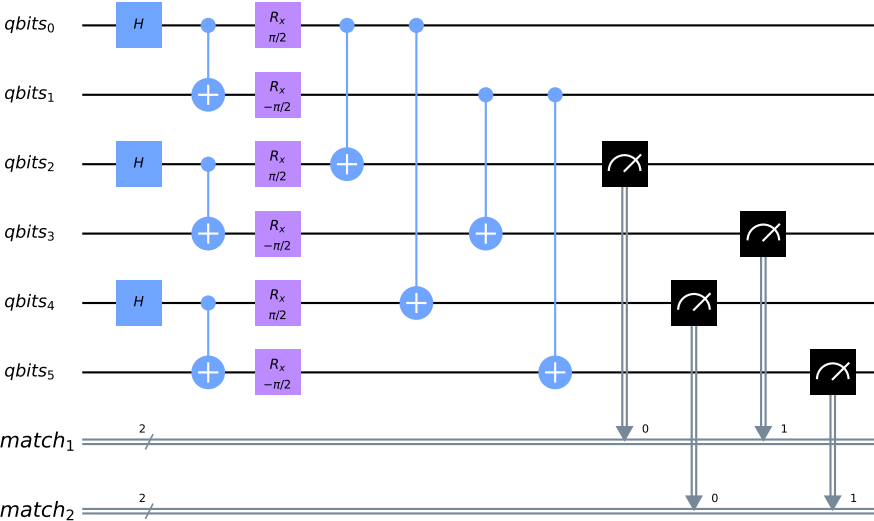}
\caption{Quantum Circuit for implementing a 3-Bell-Pair extension of Deutsch's Protocol. A Bell-Basis measurement is performed on \texttt{qbit}$_0$ and \texttt{qbit}$_1$ after performing the protocol to check the fidelity  (not shown in figure).}
\label{fig_purification_circuit}
\end{figure}

The quantum circuit for performing Bennett's Quantum Purification Protocol with Deutsch's correction operations on 3 Bell-pairs is shown in Fig. \ref{fig_purification_circuit}.

\begin{figure}[!h]
\includegraphics[scale=0.55]{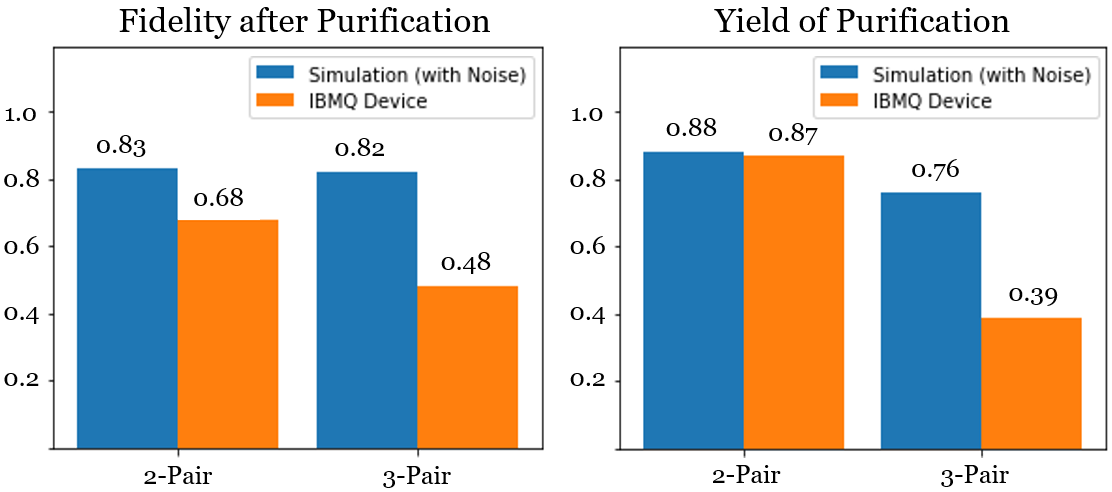}
\caption{Effect of Deutsch's Quantum Purification Protocol on (Left) Fidelity of Quantum Entanglement, and (Right) Yield of Entangled Qubits.}
\label{fig_purification_results}
\end{figure}

The results after executing this circuit on IBMQ are shown in Fig. \ref{fig_purification_results}. 

They show the effect of the number of qubits used in one round of purification on the the yield and fidelity of entanglement. We define the yield to be the percentage of times in which the classical messages between Alice and Bob agree, and, the protocol is considered successful. Since, agreement of classical messages is heralded, the fidelity of entanglement is calculated only when the protocol is successful. 

Although the 3-Pair protocol should theoretically perform better than the 2-Pair one, the results in Fig. \ref{fig_purification_results} show otherwise. This is due to the fact that, a higher number of gate-operations are required to run purification on 3 Bell-Pairs simultaneously. Since the gate operations are imperfect, this induces a greater amount of noise in the system nullifying any theoretical advantage.


\subsection{Entanglement Swapping}
\begin{figure}[!h]
\centering
\includegraphics[scale=0.75]{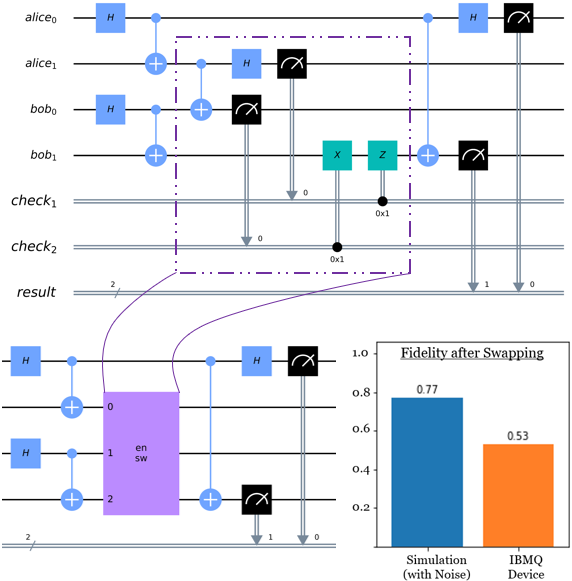}
\caption{ (Top) Quantum Circuit for performing the Entanglement-Swapping Protocol. (Bottom Left) Swapping Sub-Circuit created using Qiskit, (Bottom Right) Effect of Entanglement Swapping on the fidelity of Entanglement.}
\label{fig_enswap_res}
\end{figure}

Fig. \ref{fig_enswap_res} (Top and Bottom-Left) shows the quantum circuit for performing the Entanglement Swapping protocol with one repeater node in the middle. The results after running this circuit on IBMQ are shown on the bottom right of Fig. \ref{fig_enswap_res}.

As can be seen, entanglement swapping causes a large loss in the fidelity of entanglement. In fact, the drop in fidelity is higher than the $F^2$ scaling rule mentioned in Section 2.2. This is because, $F^2$ scaling only takes into account the role of imperfect operations during the Bell-State preparation. However, a large portion of the fidelity loss in the swapping protocol is due to the operations performed \emph{after} the state preparation and readout. A more general analysis performed by Briegel \etal \cite{briegel1998quantum} indicates this faster decrease in fidelity for highly error-prone swapping operations, which is consistent with the experimental results. Thus, further purification of the swapped qubits is necessary for another round of swapping.


\begin{figure*}[!t]
\includegraphics[scale=0.57]{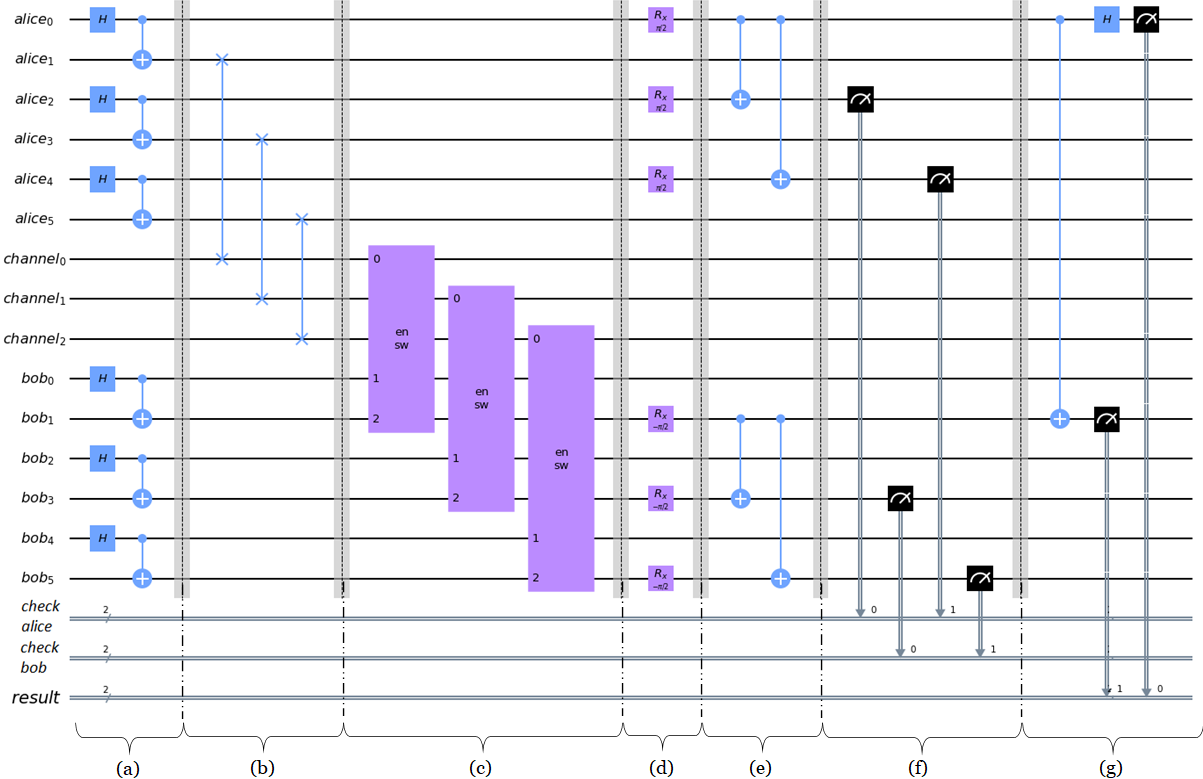}
\caption{Complete circuit of the proposed Quantum Repeater Architecture. (a) Bell-State Preparation, (b) Transmission of entangled qubits to channel, (c) Entanglement Swapping, (d) Deutsch's Correction Operation, (e) Bennett's Purification Protocol, (f) Measurement and Classical Messaging, (g) Bell-Basis Measurement to check the fidelity of entanglement.}
\label{fig_rep_final}
\end{figure*}

\subsection{Quantum Repeater}
The complete quantum circuit of a Qauntum Repeater integrating all the elements discussed earlier is shown in Fig. \ref{fig_rep_final}. 

\begin{figure}[!h]
\includegraphics[scale=0.65]{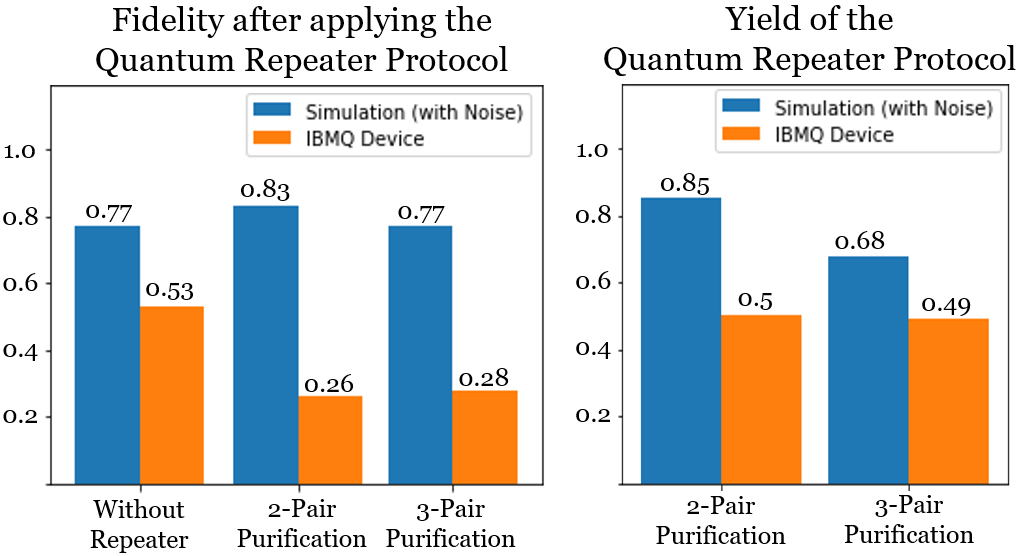}
\caption{Performance of the Quantum Repeater Circuit: (Left) Fidelity of Quantum Entanglement, and (Right) Yield of Entangled Qubits.}
\label{fig_repeater_results}
\end{figure}

In it, Alice and Bob first prepare 3 Bell-pairs. Alice transfers one qubit from each of her pairs to the channel (emulated by using SWAP gates). Then, entanglement-swapping takes place between the channel-qubits and Bob's qubits. After that, Alice and Bob use the $R_x$ gate on their qubits according to Deutsch's Protocol, and, use the CNOT operations according to Bennett's Protocol. 2 qubits each of Alice and Bob are measured out and communicated to each other to check if the Entanglement Purification was successful. Finally, The remaining qubits of Alice and Bob are measured out in the Bell-basis to check the fidelity of entanglement established between them. 

The results after running the circuit, on the final fidelity of entangled qubits and their yield, are shown in Fig. \ref{fig_repeater_results}. These demonstrate how the repeater protocol performs under hardware imperfections to establish long-range entanglement between two qubits in the existing hardware. The results are discussed further in the following section. 





\section{Discussions and Conclusion}
In this paper, we have demonstrated the complete circuit-level implementation of one of the central elements of a quantum communication network, the Quantum Repeater, and, evaluated its performance on IBM's cloud Quantum Computer. We have shown how quantum computing primitives could be used to implement a Quantum Repeater, and conversely, how the Quantum Repeater could provide a robust benchmark for state-of-the-art quantum computing hardware. 

The experiments demonstrate the efficacy of the different elements of a Quantum Repeater in increasing the fidelity of distributed Bell-Pairs and establishing long-range links. It is interesting to note that, although theoretical calculations and simulation results indicate an increase in fidelity from using a quantum repeater instead of direct transmission, the results from IBMQ indicate less stellar performance at present. To construct a full-fledged quantum repeater, a multitude of operations need to be performed on entangled qubits while protecting their coherence at the same time. However, as the number of operations increase, gate-errors, measurement-errors and dephasing become insurmountable, making the qubits unsuitable to work with. Thus, the results reveal the current state of IBM's Quantum hardware, as to the number of quantum operations that can be performed before the errors become unmitigable. Owing to this constraint, complex purification protocols with large number of operations cannot be emulated as of now. We believe that this interplay between quantum computing primitives and quantum communication protocols will help researchers to merge these two perspectives and use the techniques of each discipline to move research forward in the other. 

It is also insightful that, the QISKIT Aer simulator systematically overestimates the fidelity of the Bell-Pairs. Aer uses a simplified noise model of the device, which takes into account 4 parameters - i) Gate-Errors ii) Gate-Duration iii) $T_1, T_2$ relaxation times and iv) Readout Errors \cite{aer_noise_model}. However, a myriad of other types of errors like Cross-resonance, Static/Dynamic ZZ-Coupling, High-Order Oscillations, High-Energy Leakage, etc. relevant for the operations of an on-chip superconducting quantum processor \cite{han2020error} are not considered in the noise-model. In addition, the noise-model of QISKIT Aer is based on several calibration parameters of the IBM devices, which change from time to time. Because of these reasons, the simulation results diverge from real device behaviour. A more comprehensive noise-model is required to properly capture the effects of errors on the chip. 

There is much scope to extend this work even further. A promising direction is to come up with robust and condense Error-Correcting Codes, so that, the fidelity may be increased without discarding all the Bell-pairs. It is expected that this will result in higher yield-rates without sacrificing the error-rates. Another approach would be to co-design hardware-specific error-correcting codes, taking into account the device-architecture of a quantum system, so that errors can be mitigated more efficiently. The present work may act as a useful guide in that direction. 


\begin{acknowledgements}
We extend special gratitude to Shahnewaz Ahmed of the Dept. of Computer Science and Engineering, BRAC University for helping us to gain insight into the nature of entanglement and its role in Quantum Communication. His valuable advice, comments and suggestions are thankfully acknowledged.
 \end{acknowledgements}

\bibliographystyle{spmpsci}      

\bibliography{section/reference.bib}

\end{document}